\documentclass[aps,preprint]{revtex4}
\usepackage{amsmath}
\usepackage{amssymb}
\usepackage[dvips]{graphicx}
\begin{document}
\title{Estimation of microscopic averages from metadynamics}
\author{Guido Tiana}
\affiliation{Department of Physics, University of Milano, and INFN, via Celoria 16, 20133 Milano, Italy}
\date{\today}
\begin{abstract}
With the help of metadynamics it is possible to calculate efficiently the free energy of systems displaying high energy barriers as a function of few selected "collective variables". In doing this, the contribution of all the other degrees of freedom ("microscopic" variables) is averaged out and, thus, lost. In the following, it is shown that it is possible to calculate the thermal average of these microscopic degrees of freedom during the metadynamics, not loosing this piece of information.
\end{abstract}
\maketitle

Metadynamics is an algorithm developed by Laio and Parrinello \cite{par1,par2} to calculate the free energy of complex systems as a function of some slow--varying collective variables (CV). It consists in simulating the dynamics by adding to its energy function, at regular intervals, a non--Markovian term which depends on the collective variables and which disfavours the sampling of regions already visited. This algorithm is particularly efficient in calculating the free energy of systems displaying large energy barriers, since the non--Markovian term fills effectively the free energy wells, allowing the system to move fast to other regions of the conformational space.

In order to make the algorithm efficient, CV must be chosen in such a way that the motion of the system in the normal directions is fast and does not encounter major energy barriers. Once one obtains the free energy as a function of the CV, it is possible to calculate thermal averages for any function of the CV, just performing a weighted summation. The problem is that all information concerning variables other than CV, accounting for the microscopic motion of the system, is lost. Of course, one could first perform a metadynamic run, obtain the free energy as a function of the CV, and then perform a standard dynamics with a potential which is the sum of the system potential and the non--Markovian term. Recording the values of the microscopic variable of interest during this second run, and rescaling it to subtract the effect of the non--Markovian potential, will give its thermal average. The drawbacks of this method is not only that one has to perform two calculations, loosing all information on the microscopic variable collected in the former, but also that the latter, being a random walk in a flat energy surface, cannot distinguish between thermodynamically important and non--important regions, and consequently must sample exhaustively all parts of conformational space in order to provide the correct averages.

The idea is that, during the sampling between successive updates of the non--Markovian potential, the system can equilibrate small regions of the conformational space. The equilibrated regions usually correspond to local minima of the time--varying free energy. They can be very small, but their number very large and change during the simulation, thus covering the whole conformational space and, in particular, regions corresponding to the minima of the true free energy where the non--Markovian potential accumulates. Collecting together the averages of microscopic variables in these small regions and weighting them properly provides the correct value of the associated thermal averages.

Let $\Gamma$ be the conformational space and $\gamma\in\Gamma$ be a conformation of the system, $x(\gamma)$ a collective variable and $y(\gamma)$ a fast--varying quantity whose average one wishes to know, $V_\nu(x)$ the non--Markovian potential after $\nu$ updates, $\Delta t_D$ the time interval between two updates and $U(\gamma)$ the potential energy of the system. Assume that in the time interval from $\nu\Delta t_D$ to  $(\nu+1)\Delta t_D$ which follows the $\nu$th update, the system has sampled exhaustively a region, however small, $A_\nu\subset\Gamma$. This is the same hypothesis needed by metadynamics to work. If we indicate with square parentheses the average of any quantity over the region $A_\nu$ calculated by the metadynamics algorithm for the evolution under the total potential $U+V_\nu$, then the ergodic theorem assures that
\begin{equation}
[y]_{A_\nu}=\frac{1}{Z_\nu}\sum_{\gamma\in A_\nu}y(\gamma)e^{-\beta \{U(\gamma)+V_\nu(x(\gamma))\}},
\label{eq_erg}
\end{equation}
where $Z_\nu$ is the partition function restricted to the visited region. One can thus calculate the thermodynamic average of $y$ restricted to the visited region as
\begin{equation}
\frac{1}{Z_\nu}\sum_{\gamma\in A_\nu}y(\gamma)e^{-\beta U(\gamma)}=\left[ye^{\beta V_\nu}\right]_{A_\nu}.
\label{eq_res}
\end{equation}
The quantity $Z_\nu$ can also be calculated applying Eq. (\ref{eq_erg}) to a constant function, obtaining
\begin{equation}
\frac{1}{Z_\nu}\Omega(A_\nu)=\left[e^{\beta(U+V_\nu)}\right]_{A_\nu},
\label{eq_znu}
\end{equation}
where $\Omega(A_\nu)$ is the volume of the region $A_\nu$. From Eq. (\ref{eq_res}) one wishes to reconstruct the actual thermodynamic average of $y$, collecting together the partial averages in each region $A_\nu$ visited. In order to weight correctly the contribution coming from multiple visits of the same region of conformational space, let's partition it into regions $\Lambda_i$. Each $A_\nu$ can thus be seen as a collection of a number of regions $\Lambda_i$. The contribution of the region $\Lambda_i$ to the thermodynamic average of $y$ is
\begin{equation}
\sum_{\gamma\in \Lambda_i}y(\gamma)e^{-\beta U}=\frac{1}{n_s(\Lambda_i)}\sum_{\nu:\Lambda_i\subset A_\nu}\sum_{\gamma\in A_\nu}y(\gamma)e^{-\beta U}=\frac{1}{n_s(\Lambda_i)}\sum_{\nu:\Lambda_i\subset A_\nu}Z_\nu\left[ye^{\beta V_\nu}\right]_{A_\nu},
\end{equation}
where $n_s(\Lambda_i)$ is the number of times the system has visited region $\Lambda_i$ at different $\nu$ and $Z_\nu$ is calculated from Eq. (\ref{eq_znu}). The sum labelled by $\nu:\Lambda_i\subset A_\nu$ is meant as over all $\nu$ such that $\Lambda_i$ belongs to $A_\nu$. The average of $y$ is then the sum of the contribution of all regions $\Lambda_i$, that is
\begin{equation}
<y>=\frac{1}{Z}\sum_\gamma y(\gamma)e^{-\beta U}=\frac{1}{Z}\sum_i\frac{1}{n_s(\Lambda_i)}\sum_{\nu:\Lambda_i\subset A_\nu}Z_\nu\left[ye^{\beta V_\nu}\right]_{A_\nu}.
\label{eq_av}
\end{equation}
The partition function is found setting $y$ equal to a constant, which gives
\begin{equation}
Z=\sum_i\frac{1}{n_s(\Lambda_i)}\sum_{\nu:\Lambda_i\subset A_\nu}Z_\nu\left[e^{\beta V_\nu}\right]_{A_\nu}.
\label{eq_z}
\end{equation}
What remains to be done is to give an operative definition of the regions $\Lambda_i$. The easiest choice is to use for this purpose a bin of the collective variable $x$, and thus the sum over $i$ can be substituted by the sum over $x$.

An implementation of this algorithm is thus: 
\begin{enumerate}
\item At each step after the $\nu$th deposition, record $x$, $y$, $U$ and $V_\nu$.
\item Before the $(\nu+1)$th deposition, calculate the histogram of visited $x$. Define a threshold on frequencies  of the histogram in order to neglect the poorly visited bins. Identify the connected region of $x$ whose histogram lies above the threshold and contains the maximum, and define this set as $A_\nu$. Calculate $[\exp(\beta V_\nu)]$, $[y \exp(\beta V_\nu)]$ and $[\exp(-\beta (U+V_\nu))]$ over $A_\nu$. Calculate $Z_\nu$ from Eq. (\ref{eq_znu}) assuming that $\Omega(A_\nu)$ is proportional to the range of $x$ spanned.
\item At the end of the simulation, calculate $n_s(\Lambda_i)$ from the recorded $A_\nu$. Then calculate the total partition function from Eq. (\ref{eq_z}) and the average $<y>$ from Eq. (\ref{eq_av}).
\end{enumerate}

\begin{figure}
\includegraphics[height=6cm]{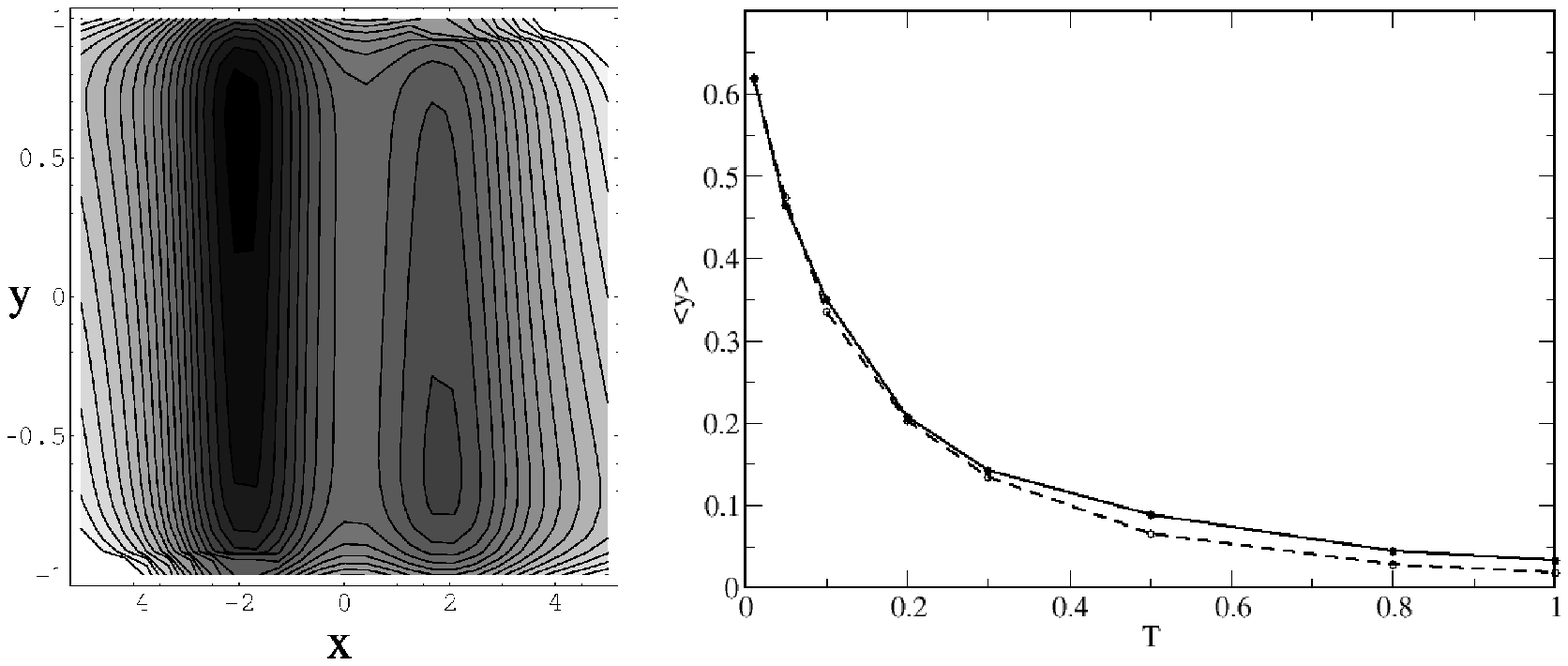}
\caption{(left) The potential energy of the system as described by Eq \protect\ref{equ}. Each level of the contour plot corresponds to 0.2 energy units. (right) The thermal average $<y>$ as a function of temperature obtained from metadynamic calculations (solid curve) and integrated numerically (dashed curve).}
\label{fig1}
\end{figure}

We have tested the above idea on a simple, two--dimensional system controlled by the energy function
\begin{equation}
U(x,y)=-2\exp[-(x+2)^2/2]-\exp[-(x-2)^2/2]+x^2/20+xy/10+\exp(y^5)+\exp(-y^5),
\label{equ}
\end{equation}
which displays two wells, separated by a barrier along the $x$--direction (see Fig. \ref{fig1}) whose height is of the order of unity, in the arbitrary energy units defined by Eq. (\ref{equ}). Within each well, the system does not display any barrier, complying with the requirement that $y$ is fast--varying. The system is constrained to the region $-10<x<10$, $-1<y<1$.

Performing metadynamic calculations employing $x$ as collective variable and lasting for $10^7$ steps (with deposition time $t_D=50000$ steps (i.e., very long, the dependence of the equilibration ability on $t_D$ will be studied in the following), height of the Gaussian functions equal to $0.005$, standard deviation $0.01$, binning of $x$ $0.01$, threshold on the histogram $10^{-3}$, diffusion coefficient $2.5\cdot 10^{-3}$ step$^{-1}$), one obtains the thermal average $<y>$ of the "microscopic" quantity $y$. The value of $<y>$ obtained from this calculation is displayed in Fig. \ref{fig1}  as a function of the temperature, together with the actual value of $<y>$ calculated as $Z^{-1}\int dx\,dy\;y\,\exp(-U(x,y)/T)$, where $Z$ is the normalization factor. The agreement is good at low temperatures and worsen at high temperatures. This is not unexpected, since metadynamics is designed to overcome large energy barriers (i.e., $\gg T$), not to speed up the sampling of flat free--energy surfaces.

\begin{figure}
\includegraphics[height=6cm]{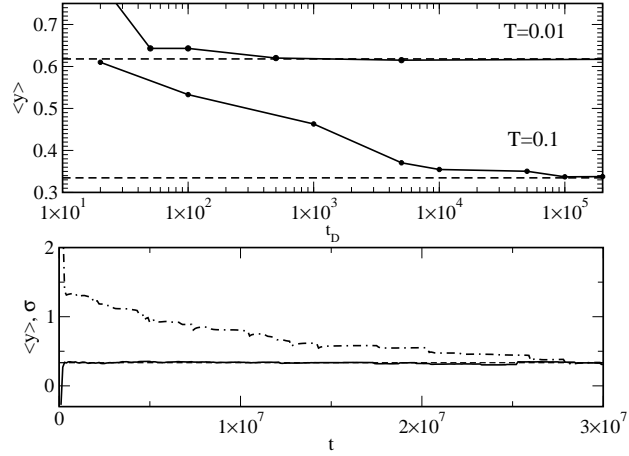}
\caption{(upper panel) The thermal average $<y>$ at $T=0.1$ (lower solid curve) and $T=0.01$ (upper solid curve) as a function of the deposition time $t_D$. The dashed line indicate their respective actual value. (lower panel) The thermal average $<y>$ (solid curve) and the standard error $\sigma$ of the calculated free energy with respect to the actual one, as functions of the simulation length $t$ at $T=0.1$.}
\label{fig2}
\end{figure}

The value of $t_D$ used above is very large, in order to check the correctness of the algorithm in the conditions under which Eq. (\ref{eq_erg}) certainly holds. To be computationally efficient, $t_D$ has to be as small as possible \cite{par2}. In the upper panel of Fig. \ref{fig2} is displayed the value of $<y>$ at $T=0.1$ and $T=0.01$, calculated from simulations performed with different update times $t_D$. At both temperatures the average of $y$ reaches its true value for large $t_D$. For low $t_D$, $<y>$ is largely overstimated, as a result of the fact that the system is not able to equilibrate regions along the y--direction, and thus the algorithm fails. The threshold of $t_D$ needed for such an equilibration seems to be strongly dependent on the temperature. While at $T=0.01$ a deposition time of $100$ is enough to obtain a $<y>$ with a 5\% error, at $T=0.1$ one needs a $t_D$ of the order of $10^4$.

In the lower panel of Fig. \ref{fig2} is displayed the time needed by $<y>$ and by the free energy $F(x)$ to reach their true values at $T=0.1$. While $<y>$ is correctly computed in $10^6$ steps, the free energy converges to the equilibrium one, given by $F(x)=-T\log\int dy\;\exp(-U(x,y)/T)$ in $10^7$ steps, indicating that the thermal average of the microscopic variable is correclty calculated during the sampling, and not after that the metadynamics has flattened the free energy surface. The fast convergence of $<y>$ with respect to $F(x)$ is associated with the fact that the value of $<y>$ is determined mainly by small, low free energy regions of conformational space, while $\sigma$ indicates the convergence of the whole $F(x)$.

Summing up, the values assumed by microscopic variables during a metadynamic run can be collected and weighted properly in order to obtain their thermal average at no further cost. The hypotheses under which the algorithm works are that $V_\nu$ converges to the true free energy of the system, that it is possible to label regions where the system equilibrates (i.e., to define non--empty $A_\nu$) and that it is possible to evaluate the volume $\Omega(A_\nu)$ of such regions. 

\begin{figure}
\includegraphics[height=6cm]{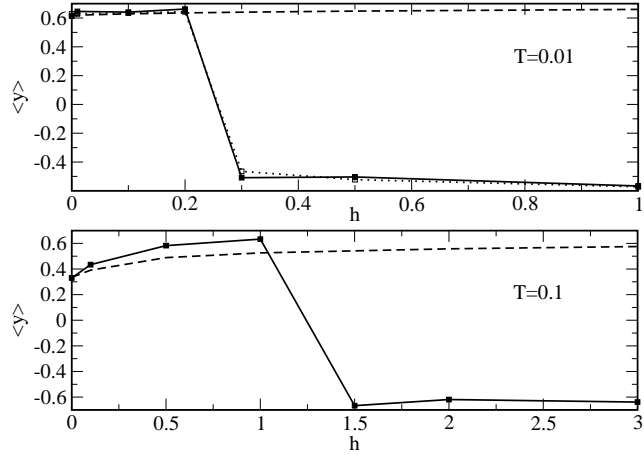}
\caption{The values of $<y>$ calculated placing barriers of different heights $h$ along the $y$--coordinate. The dashed curve indicates the correct value. (above) The vaule of $<y>$ obtained at $T=0.01$ using $t_D=200$ (solid curve) and $t_D=2000$ (dotted curve). (below) The same at $T=0.1$ and $t_D=10^4$. The simulations lasted for $10^4 t_D$. }
\label{fig3}
\end{figure}

In order to test the effects of roughness in the energy landscape along the $y$ direction, we have added to the potential of Eq. (\ref{equ}) a term
\begin{equation}
h\cdot\exp(-\frac{y^2}{2\cdot 0.2^2}),
\end{equation}
where $h$ is an energy parameter which we can be tuned. In Fig. \ref{fig3} it is shown the behaviour of the predicted $<y>$ with respect to $h$. At the lower temperature $T=0.01$ and using deposition times $t_D=200$ or $t_D=2000$, the correct value of the average is obtained up to $h=0.2$. Above this value, the system is no longer able to diffuse along the $y$ directions and thus the resulting average is incorrect. At $T=0.1$, using $t_D=10^4$, the algorithm is able to calculate a reasonable value of $<y>$ up to $h=1$, although the associated error ranges from 5\% to 20\% as $h$ is increased from 0.1 to 1. 

\begin{figure}
\includegraphics[height=6cm]{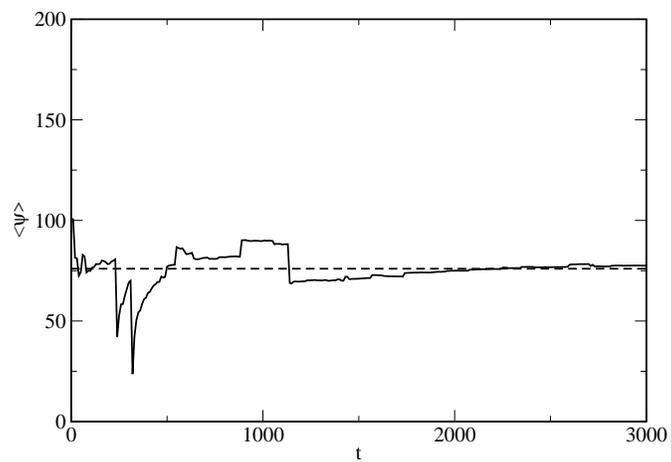}
\caption{The values of $<\psi>$ of dialanine, calculated with the above algorithm, as a function of the number of updates of the non--Markovian term (each of 0.6 ps). The dashed line is the true value.}
\label{fig4}
\end{figure}

The case discussed above is a simple example where the averages can be calculated analyticaly, meant to illustrate the algorithm. A further test has been performed on a more realistic system, that is dialanine, which has been widely characterized \cite{brand} in terms of the Ramachandran dihedrals $\phi$ and $\psi$. In particular, it has been shown to display at $300$K a barrier of several $kT$ along the $\phi$ direction, while it is smoother along the $\psi$ direction. The thermal average $<\psi>$ calculated through a metadynamic run using both $\phi$ and $\psi$ as CV gives $76^o$. Performing a metadynamics using as CV only $\phi$ (with updating time 0.6 ps, height of the Gaussians equal to $0.1$ kcal/mol) gives the result reported in Fig. \ref{fig4}, indicating that a good estimate of $\phi$ can be found after approximately $1500$ updates of the non--Markovian term.

We have thus shown that from a metadynamics run it is possible to extract the thermal averages of micorscopic quantities, under the same hypotheses which allow metadynamics to work.

\begin{acknowledgments}
As usual, I have to thank Max Bonomi, Davide Provasi and Ludovico Sutto for deep discussions and help. I acknowledge the financial support of the 2003 FIRB program of the Italian Ministry for Scientific Research.
\end{acknowledgments}

\clearpage
\newpage
\clearpage
\newpage

\clearpage
\newpage

\clearpage
\newpage

\end{document}